\documentclass[review]{elsarticle}

\pdfoutput=1

\usepackage{hyperref}
\usepackage[anythingbreaks]{breakurl}
\usepackage{mathtools}
\usepackage{graphicx}
\usepackage[table,xcdraw]{xcolor}
\usepackage{subcaption}
\usepackage{CJKutf8}
\usepackage[bottom]{footmisc}
\usepackage{siunitx}

\usepackage{textcomp}
\usepackage{textgreek}

\renewcommand{\vec}[1]{\mathbf{#1}}

\journal{The 2018 International Conference on Artificial Life and Robotics (ICAROB2018)}

\begin{document}

\begin{frontmatter}

\title{Emotional Contribution Analysis of Online Reviews}


\author[gidai]{Elisa Claire Alem\'an Carre\'on
\corref{mycorrespondingauthor}}
\ead{s153400@stn.nagaokaut.ac.jp}

\author[gidai]{Hirofumi Nonaka}
\ead{nonaka@kjs.nagaokaut.ac.jp}

\author[nagasaki]{Toru Hiraoka}
\ead{hiraoka@sun.ac.jp}

\author[miyazaki]{Minoru Kumano}
\ead{kumano@cc.miyazaki-u.ac.jp}

\author[okayama]{Masaharu Hirota}
\ead{hirota@mis.ous.ac.jp}

\author[hiroshima]{Takao Ito}
\ead{itotakao@hiroshima-u.ac.jp}

\address[gidai]{Nagaoka University of Technology, Nagaoka, Japan}
\address[miyazaki]{University of Miyazaki, Miyazaki, Japan}
\address[nagasaki]{University of Nagasaki, Nagasaki, Japan}
\address[okayama]{Okayama University of Science, Okayama, Japan}
\address[hiroshima]{Hiroshima University, Hiroshima, Japan}

\cortext[mycorrespondingauthor]{Corresponding author}

\begin{abstract}

In response to the constant increase in population and tourism worldwide, there is a need for the development of cross-language market research tools that are more cost and time effective than surveys or interviews. Focusing on the Chinese tourism boom and the hotel industry in Japan, we extracted the most influential keywords in emotional judgement from Chinese online reviews of Japanese hotels in the portal site \textit{Ctrip}. Using an entropy based mathematical model and a machine learning algorithm, we determined the words that most closely represent the demands and emotions of this customer base. 

\end{abstract}


\end{frontmatter}

\section{Introduction}\label{intro}

Worldwide, the population of Chinese tourists has recently increased significantly over the years. This increase has led to a significant impact to several industries over the world, most directly the hotel industry, as their customer bases changed. In addition to this impact on business, there is a current increase of academic research across the world about Chinese tourist populations\cite[][]{sun2017}. Now, there was a particularly notable increase in the Chinese tourist population in Japan of 107.3\% from 2014 to 2015, and a total number of tourists of \num[group-separator={,}]{6372948} in 2016 (Japan National Tourism Organization, 2017). However, while there are studies of Chinese tourists in several countries, for example, one researching the satisfaction levels of Chinese tourists in Vietnam\cite[][]{truong2009} or another studying their food preferences in Australia\cite[][]{chang2010}, these studies are done mainly outside Japan, which has been largely impacted by Chinese tourism. Many if not most of these studies are performed based on the results of surveys or interviews, as are the market studies performed by companies. However, depending on the formulation of the questions, the results of questionnaire based surveys can be altered and they also present difficulties and costs of the survey by itself and in this case, the need for translations. Furthermore, the sample size of these survey studies is commonly limited to tens or hundreds of people at most. On the other hand, Data mining and text mining techniques can not only cheaply and quickly gather tens of thousands if not more samples, but the source of this extensive amount of information also can be thought to be unaltered by any part of the data extraction process. Considering this, in our study, we developed a text-mining method to analyze the demands and needs of Chinese customers of Japanese Hotels, exploring our results in the perspective of business or management decisions. In order to perform this analysis, we have extracted a large quantity of text reviews from a Chinese portal site \textit{Ctrip} (http:// http://www.ctrip.com), and determined the most commonly used words that would contribute the most to emotional judgement behind positive and negative opinions in a review using an entropy based mathematical extraction method. These extracted keywords related to emotional judgement not only allow us to perform a Support Vector Machine based emotional classification of the reviews, but conceptual words in these lists bring insight into which concrete topics are the Chinese tourists concerned with. After classifying the sentences in the extracted reviews as positive and negative with an optimized SVM, we have analyzed the weight value assigned to them by the SVM. For words that are not part of the support vector this is equal to 0; however, the support vector lets us observe the words that, while potentially close to the border between positive and negative sentiments, provide a strong and clear distinction in the emotional classification. These words allow us to analyze the writing tendencies of Chinese customers in either positive or negative reviews. We also observed the frequency of the terms in all of the reviews to extract the most utilized words in either kind of reviews. This is valuable information for making business management decisions from the point of view of the hotel industry companies. Because these keywords are most related to customer satisfaction, a company can improve customer service or facilities to increase profit. 

\section{Methodology}

\subsection{Data collection and processing}

In this study, we used the HTML parsing python library \textit{BeautifulSoup}, and the local database management tool SQLite  using the python library \textit{sqlite3} to automatize data managing processes. Different from the English language, Chinese texts lack a separation between different words, and as such, when collecting these texts, they are all a single string of characters. To be able to perform a statistical analysis of each word, the Stanford Word Segmenter\cite[][]{chang2008} program developed by the Stanford NLP Group was implemented for this task using the python \textit{nltk} library. During the segmentation of all the words in our corpus, irregularities occurred where the reviews were written in other languages or where unusual punctuation marks were used. We designated a list of characters that could be recognized as noise and then cleared all the text in the corpus of these characters.

\subsection{Sentiment analysis}

In order to determine the words that clearly impact the user’s emotional judgement we calculated the entropy value of each word in relation to each class. Shannon’s Entropy, in the field of Information Theory, is defined to be the expected value of the information content in a signal\cite[][]{shannon1948} or it can be thought of as the grade of impossibility of predicting an outcome. Using this value, we can observe the probability distribution of each word inside the corpus. A word that is included in many documents, it will have a high entropy value for that set of documents, since it becomes uncertain to predict in which document it will appear. Opposite to this, a word appearing in only one document will have an entropy value of zero, since it is completely predictable. We show this concept in the figure below. To apply this logic, we retrieved a sample of our corpus and with the collaboration of a group of Chinese students, tagged each sentence as the classes positive or negative depending on the emotion that the text conveyed, then calculated the entropy values for each word in relation to the set of sentences from each class. Words with higher entropy relating to the positive set than to the negative set by a factor of \(\alpha\) were determined to be keywords influencing positive emotional judgement in Chinese reviews of hotels. Likewise, words with higher entropy for the negative set than the positive set by a factor of \(\alpha'\) were determined to be keywords related to the negative emotional judgement in our texts. 

Support Vector Machines are supervised machine learning models usually applied to classification or regression problems\cite[][]{cortes1995}. We use it to classify the rest of our corpus into emotional classes in our study. An SVM is trained to classify data based on previously labeled data, generalizing features of the data by defining a separating (p-1)-dimensional hyperplane in a p-dimensional space in which each dimension is a feature of the data. The separating hyperplane, along with the support vectors, divide the multi-dimensional space and minimize the error of classification. In our study we used the linear kernel of the SVM classification process. Each training sentence is a point of data, a row in the vector \(\vec{x}\), where each column represents a feature, in our case the quantities of each of the keywords in that particular sentence. The labels of previously known classifications (1 for positive, 0 for negative) for each sentence comprise the \(f(x)\) vector. The Weight Vector \(\vec{w}\) is comprised of the influences each point has had in the training process to define the angle of the hyperplane and the bias coefficient \(b\) determines its position. In the field of corpus study and Natural Language Processing, each of the features of a data point is the number of times that a word is included in a document. In our study we used a number of keyword lists, defined by our entropy calculations with different comparison coefficients, as the possible features; trained the SVMs implementing the Support Vector Classifier included in the python library \textit{scikit-learn}; managed the vector mathematics with the mathematical python library \textit{numpy}; and tested for each one using the K-fold Cross Validation method, which has been proven to provide good results in small samples\cite[][]{kohavi1995}. In each test we calculated the Precision, Recall and \(F_1\)-measure\cite[][]{powers2011} for our predictions. As stated earlier in section, each point of data that is classified incorrectly causes a change in the weight vector to better locate the separating hyperplane and classify new data correctly. These changes to the weight vector are strong for features that needed to be taken account of to classify with a minimal error, those contained in the support vectors, close to the separating hyperplane. Sequentially, the weight vector can be interpreted as a numerical representation of the effect each feature, or in our case, each of these normally ambiguous words, had for the classification process and the class it has decisive influence in. Because the weight vector assigns value to the words that comprise the support vectors, the words with higher weights are thought to be closer to the dividing hyperplane, while still clearly and decisively belonging to one of the categories the hyperplane divides the high-dimensional space in.

\section{Results and Discussions}

After having our training data tagged by a group of Chinese student collaborators, we experimented with different comparison coefficients for the entropy values calculated from ‘positive’ and ‘negative’ emotional classes. The mutually independent coefficients \(\alpha\) and \(\alpha'\) were tested from 1.5 to 3.75 in intervals of 0.25. The result was 20 lists, 10 for each emotional class.  

At the beginning we experimented with different kernels for the SVM, as well as some Ensemble Learning methods, like the Boosting, Voting and Stacking. We ultimately decided to use the linear kernel for the benefits of the weight vector obtainability. We also experimented with different parameters for the SVC, finding that the best performing value for \(C\), a constant that affects the optimization process when minimizing the error of the separating hyperplane. Low values give some freedom of error minimizes false positives, but depending on the data it can increase false negatives. Inversely, high values of \(C\) will likely result in minimal false negatives, but a possibility of false positives. We found the ideal value for this parameter was \(C=0.5\) in our final classifier. We trained a different Support Vector Classifier with each of the 20 lists, and we chose the best performing lists for each emotional class, resulting in a positive emotion classifier (positive or not positive), based on the results of a k-fold cross validation process in which we calculated their accuracy and F-measure means and standard deviations. Table \ref{tab:1} shows the results of our system performance. After observing the misclassification behavior for the positive emotion classifier, which mostly misclassified negative sentences, we decided to combine both keyword lists into a single large list to train the positive emotion classifier. 

Below Table \ref{tab:2} we show some of our keywords from the subject class (words that help us understand demands of the users) that have a relatively high weight value for both positive and negative extremes, and their translations in the relevant context.

\begin{table}[]
\centering
\caption{Performance of our method.}
\label{tab:1}
\begin{tabular}{|l|l|l|l|l|l|} 
\hline
Emotion & \(C\) & Accuracy \(\mu\) & Accuracy \(\sigma\) & \(F_{1}\) \(\mu\) & \(F_{1}\) \( \sigma\) \\ \hline
Positive (\(\alpha=2.75\)) & 0.5 & 0.88 & 0.15 & 0.90 & 0.09 \\ \hline
\end{tabular}
\end{table}

Analyzing the extracted subject keywords is the key to performing a market study of Chinese customers of Japanese Hotels. They can be interpreted as explicit needs and demands. 

Analyzing words with an influence of positive emotions, we found that the most relevant subjects Chinese customers perceive positively are cleanliness and size, very possibly of the room they had stayed in. There is also the possibility that reviewers were praising in general the cleanliness of Japan’s environment, easily accessible and their culture of respecting spaces.

\begin{table}[]
\centering
\caption{Positive Emotion Keywords.}
\label{tab:2}
\begin{tabular}{|l|l|l|}
\hline
Word & English & Weight \\ \hline 
\begin{CJK}{UTF8}{gbsn} 干净 \end{CJK} & Clean & 0.638 \\ \hline
\begin{CJK}{UTF8}{gbsn} 大 \end{CJK} & big, wide & 0.624 \\ \hline
\begin{CJK}{UTF8}{gbsn} 交通 \end{CJK}  & Traffic & 0.586 \\ \hline
\end{tabular}
\end{table}

\section{Conclusion}

In our study, with the purpose to understand emotional responses of Chinese customers of Japanese hotels, their demands and needs, we extracted keywords from their reviews from the Chinese portal site Ctrip using entropy calculations from a manually classified sample of our data; then we used these keywords in machine learning experiments. Using the keywords to train a linear kernel Support Vector Classifier, we obtained the highest performance entropy based keywords.
Using the weight vectors of our classifiers, as well as frequency of the words in our data set, we found that Chinese customers have a preference for big and clean rooms, big thermal baths or bathhouses, expect good cost-performance regardless of price, that there is a lack of Chinese text translation and that they prefer hotels where breakfast is included. 
In future works we plan to investigate further into this topic, extending our data set, researching for other factors, such as time, scoring, revenue, and others. Another subject to investigate is to determine subjects that are classifiable as general topics in tourism compared to those only applicable in hotels. As mentioned before, we would also like to study the food preferences of Chinese tourists. Additionally, it would be interesting to study further into more specific emotions than the positive and negative classification we performed in our study.

\section*{Acknowledgements}

This research is supported by Japan Construction Information Center Foundation (JACIC). 

\section*{References}

\bibliography{icarob}

\end{document}